\documentclass[12pt]{article}
\usepackage{amsfonts,amsmath,amssymb,graphicx}
\usepackage{ifpdf}
\ifpdf
\usepackage{epstopdf,hyperref}
\else
\usepackage[hypertex]{hyperref}
\fi

\advance\textheight by \topskip
\usepackage[all]{xy}

\begin{document}

\thispagestyle{empty}
\def\thefootnote{\fnsymbol{footnote}}
\begin{flushright}\sf\footnotesize
KCL-MTH-09-10\\
ZMP-HH/09-20\\
Hamburger Beitr\"age zur Mathematik Nr.\,348
\end{flushright}

\vskip 2.5em

\begin{center}\LARGE
Defect lines, dualities, and generalised orbifolds
\end{center}\vskip 2em

\begin{center}
J\"urg Fr\"ohlich$\,{}^1$, J\"urgen Fuchs$\,{}^2$, Ingo Runkel$\,{}^{3,4}$, Christoph Schweigert$\,{}^4$
\end{center}

\vskip 1em

\begin{center}
${}^1$ Theoretische Physik, ETH Z\"urich,\\
8093 Z\"urich, Switzerland
\end{center}

\begin{center}
${}^2$ Teoretisk fysik, Karlstads Universitet,\\ 
Universitetsgatan 21, 65188 Karlstad, Sweden
\end{center}

\begin{center}
${}^3$ Department of Mathematics, King's College London,\\
Strand, London WC2R 2LS, United Kingdom
\end{center}

\begin{center}
${}^4$ Department Mathematik, Universit\"at Hamburg,\\
Bundesstra\ss e 55, 20146 Hamburg, Germany
\end{center}

\vskip 2.5em

\begin{center}\footnotesize\emph{
Contribution to the proceedings of the\\ 
XVI International Congress on Mathematical Physics, Prague, 3.8\,--\,8.8.2009
}\end{center}
\vskip 2em

\begin{abstract}
Defects are a useful tool in the study of quantum field theories. This is illustrated in the example of two-dimensional conformal field theories. We describe how defect lines and their junction points appear in the description of symmetries and order-disorder dualities, as well as in the orbifold construction and a generalisation thereof that covers exceptional modular invariants.
\end{abstract}

\setcounter{footnote}{0}
\def\thefootnote{\arabic{footnote}}

\newpage

\section{Introduction}

Defects in a quantum field theory or in a model of statistical mechanics correspond to `inhomogeneities' in correlators localised on hypersurfaces. Expectation values of fields can have discontinuities or singularities at such hypersurfaces. Defects appear naturally in connection with duality symmetries. For example, in electric-magnetic duality of four-dimensional supersymmetric Yang-Mills theory \cite{Montonen:1977sn,Kapustin:2006pk}, Wilson loop operators get exchanged with {}'t~Hooft operators, and the latter prescribe a singularity of the gauge field along a line. Another example, more in line with the present exposition, is Kramers-Wannier duality of the two-dimensional Ising model on a square lattice \cite{Kramers:1941kn,Savit:1979ny}. In that context
the correlator of Ising spins at lattice sites $i$ and $j$ at a given temperature is equal to the (suitably normalised) partition function of the Ising model on a lattice with a line of frustration running from $i$ to $j$, evaluated at the dual temperature. Across the line of frustration, the sign of the interaction of neighbouring spins is inverted.

We will concentrate on one-dimensional defects in two-dimensional conformal field theory (CFT). These describe universality classes of defect lines in lattice models, but they are also a useful tool for investigating properties of the CFT itself. At first glance, it seems that defect lines will not give more insight than the study of boundary conditions, because by `folding' the surface back onto itself along a defect line, the defect becomes a boundary of the folded model \cite{Wong:1994pa}. However, defects have one important property not shared by boundaries, and which cannot be captured by the `folding trick':
an arbitrary number of defect lines can join at {\em defect junctions}. Below we will outline
two applications in which defect junctions are important. One is the description of order-disorder dualities via defects, and the other is the construction of orbifold models.

\section{Symmetries and order-disorder dualities}

At a defect line on the world sheet one must specify boundary conditions for the fields on either side of the defect; 
we refer to these as {\em defect conditions}. To a defect with defect condition $X$ one can assign an operator $D_X$ on the space $\mathcal{H}$ of bulk states of the CFT. 
This operator can be defined
by placing a state $\phi \,{\in}\, \mathcal{H}$ at zero and the defect $X$ on the unit circle; the resulting state is $D_X \phi$. There are two compatibility conditions between a defect $X$ and the conformal symmetry. Denoting the modes of the holomorphic and anti-holomorphic component of the stress tensor by $L_m$ and $\bar L_m$, we call $X$ {\em conformal\/} iff
$[L_m-\bar L_{-m},D_X] = 0$ for all $m \,{\in}\, \mathbb{Z}$, and we call it {\em topological\/} iff it satisfies the stronger condition $[L_m,D_X] = 0 = [\bar L_m,D_X]$ for all $m \,{\in}\, \mathbb{Z}$. Thus a defect is conformal iff the corresponding boundary condition in the folded model is conformal, and it is topological iff it is transparent to all components of the stress tensor. The qualifier `topological' derives from the fact that such defects are tensionless and can be deformed on the world sheet without affecting the value of correlation functions, as long as the defect line is not taken across field insertions or other defect lines.

Finding all conformal defects for a given CFT is a difficult problem. The only CFTs for which all conformal defects are known are the Lee-Yang model of central charge $-\tfrac{22}{5}$ and the Ising model\,\footnote{~Strictly speaking, only conformal defects that have a discrete content of defect fields are known
\cite[Sect.\,4.1]{Quella:2006de}.}
of central charge $\tfrac12$ \cite{Oshikawa:1996dj,Quella:2006de}. Even for a single free boson, only a subset of conformal defects is known \cite{Bachas:2001vj}. Topological defects preserve more symmetry, and for these
the situation is better: one knows all topological defects for the Virasoro minimal models \cite{Petkova:2000ip} and for the free boson \cite{Fuchs:2007tx}. 

Consider, for example, the Ising model. It has three elementary topological defects, labelled ${\bf 1}$, $\sigma$, 
$\varepsilon$; all other topological defects are superpositions of these. The defect ${\bf 1}$ is just the trivial defect, and its defect operator $D_{\bf 1}$ is the identity. The two other defect operators have the composition rules $D_\varepsilon D_\varepsilon = D_{\bf 1}$, $D_\varepsilon D_\sigma = D_\sigma D_\varepsilon = D_\sigma$, and $D_\sigma D_\sigma = D_{\bf 1} + D_\varepsilon$.
The composition of defect operators for defects $X$ and $Y$ can be realised by
placing the defect $Y$ on the unit circle, the defect $X$ on a circle with radius $r\,{>}\,1$ and taking the limit $r \,{\rightarrow}\, 1$. For topological defects, this limit is well-defined (the correlator is independent of $r$), and in the limit one obtains a new defect on the unit circle, called the {\em fused defect}, whose defect operator is $D_X D_Y$. In this way one obtains the {\em fusion algebra of topological defects} \cite{Petkova:2000ip,tft1}. Note that, for the Ising model, this fusion algebra coincides with the fusion rules of Virasoro highest weight representations;
this also holds for all A-type Virasoro minimal models \cite{Petkova:2000ip}.

We can easily interpret $D_\varepsilon$: it implements the $\mathbb{Z}_2$-symmetry of the Ising model given by inverting the sign of the spin field. The $\sigma$-defect is related to oder-disorder duality \cite{Frohlich:2004ef,Frohlich:2006ch}. To see this we make use of the following rules for taking the $\sigma$-defect past field insertions:
\begin{equation} \label{eq:sig-defect-past-field}
\setlength{\unitlength}{1.2pt}
\raisebox{-20pt}{\begin{picture}(270,45)(0,0)
  \put(0,0){
    \put(13,0){
      \put(0,0){\scalebox{1.2}{\includegraphics{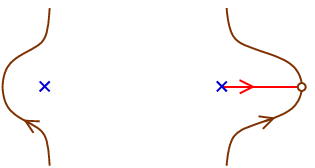}}}
      \put(15,38){\scriptsize $\sigma$ }
      \put(72,36){\scriptsize $\sigma$ }
      \put(75,26){\scriptsize $\varepsilon$ }
      \put(9,16){\scriptsize $\sigma(z)$ }
      \put(52,14){\scriptsize $\mu(z)$ }
    }
    \put(47,23){ $=$ }
    }
    \put(160,0){ \put(-35,21){and}
    \put(-2,0){
      \put(0,0){\scalebox{1.2}{\includegraphics{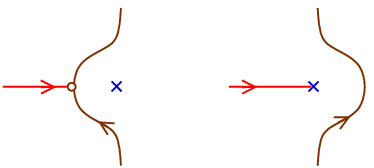}}}
      \put(5,18){\scriptsize $\varepsilon$ }
      \put(74,18){\scriptsize $\varepsilon$ }
      \put(37,39){\scriptsize $\sigma$ }
      \put(95,39){\scriptsize $\sigma$ }
      \put(32,17){\scriptsize $\sigma(z)$ }
      \put(84,16){\scriptsize $\mu(z)$ }
    }
    \put(47,23){ $=$ }
    }
\setlength{\unitlength}{1pt}
\end{picture}}
\end{equation}
The first equality says that taking a $\sigma$-defect past a spin field $\sigma(z)$
produces a disorder field $\mu(z)$, i.e.\ a `1-fold junction field' emitting an $\varepsilon$ defect line. 
The second equality says that, conversely, a disorder field gets turned into a spin field. Note that to formulate these rules, 1-fold and 3-fold defect junctions are needed. While $\mu(z)$, like $\sigma(z)$, has 
left/right conformal weight $(\tfrac1{16},\tfrac1{16})$, 
the field inserted at the 3-fold junction is a {\em topological junction field}, i.e.\ a junction field for which the correlators do not depend on the precise location of the junction, so that it can be moved on the world sheet.
For this to be the case the junction field must have conformal weight $(0,0)$. That the equalities \eqref{eq:sig-defect-past-field} hold inside correlators is not obvious, but can be proved within the three-dimensional topological field theory approach to rational CFT \cite{tft1,Frohlich:2004ef,Frohlich:2006ch}. With these rules one obtains the following series of identities for correlators on the sphere:
\begin{equation*}
\setlength{\unitlength}{1.1pt}
\raisebox{-24pt}{\begin{picture}(84,55)(0,0)
    \put(0,0){
      \put(0,0){\scalebox{1.1}{\includegraphics{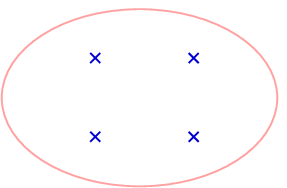}}}
      \put(0,-63){
      \put(25,104){\scriptsize $\sigma$ }
      \put(25, 71){\scriptsize $\sigma$ }
      \put(54,104){\scriptsize $\sigma$ }
      \put(54, 71){\scriptsize $\sigma$ } }
    }
\end{picture}}
  \,{=}\, \frac{1}{\sqrt{2}}\,
\raisebox{-24pt}{\begin{picture}(84,55)(0,0)
    \put(0,0){
      \put(0,0){\scalebox{1.1}{\includegraphics{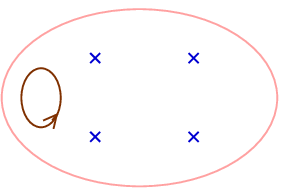}}}
      \put(0,-63){
      \put(19,87){\scriptsize $\sigma$ }
      \put(25,104){\scriptsize $\sigma$ }
      \put(25, 71){\scriptsize $\sigma$ }
      \put(54,104){\scriptsize $\sigma$ }
      \put(54, 71){\scriptsize $\sigma$ } }
    }
\end{picture}}
    \,{=}\, \frac{1}{\sqrt{2}}\,
\raisebox{-24pt}{\begin{picture}(84,55)(0,0)
    \put(0,0){
      \put(0,0){\scalebox{1.1}{\includegraphics{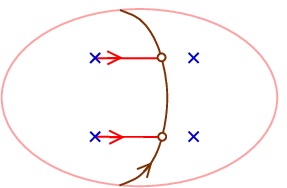}}}
   \put(-39,0){
    \put( 89,24){\scriptsize $\sigma$ }
    \put( 75,32){\scriptsize $\varepsilon$ }
    \put( 75,17){\scriptsize $\varepsilon$ } }
      \put(0,-63){
      \put(25,104){\scriptsize $\mu$ }
      \put(25, 71){\scriptsize $\mu$ }
      \put(54,104){\scriptsize $\sigma$ }
      \put(54, 71){\scriptsize $\sigma$ }}
    }
\end{picture}}
=
\raisebox{-24pt}{\begin{picture}(84,55)(0,0)
    \put(0,0){
      \put(0,0){\scalebox{1.1}{\includegraphics{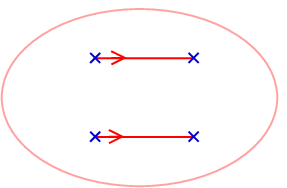}}}
   \put(-145,0){
    \put(187,32){\scriptsize $\varepsilon$ }
    \put(187,17){\scriptsize $\varepsilon$ } }
      \put(0,-63){
      \put(25,104){\scriptsize $\mu$ }
      \put(25, 71){\scriptsize $\mu$ }
      \put(54,104){\scriptsize $\mu$ }
      \put(54, 71){\scriptsize $\mu$ }}
    }
\setlength{\unitlength}{1pt}
\end{picture}}
\end{equation*}
In the first step, a small circle of $\sigma$-defect is inserted, 
which modifies the correlator by a factor of $\sqrt{2}$; then we increase the size of the defect circle, taking it past the field insertions, and in the end collapse it again, thereby cancelling the factor $\sqrt{2}$. Altogether, it follows that the correlator of four spin fields is equal to that of four disorder fields, which is an archetypical example of order-disorder duality. 

This reasoning generalises to more complicated correlators and to arbitrary rational CFTs. The fusion algebra of defect lines preserving the rational symmetry contains a subset of elementary defects, called {\em group-like defects}. They are characterised by the composition rule $D_X \bar D_X = \mathrm{id}$, where $\bar D$ denotes the defect operator for the defect line with reversed orientation. These defects form a group which acts as symmetries of the correlators. Order-disorder dualities can be read off from the fusion algebra as well \cite{Frohlich:2006ch}:

\medskip\noindent\emph{If, for a defect $X$, the composition $D_X \bar D_X$ is a sum of only group-like defect operators, then $X$ describes an order-disorder duality of the CFT.}

\medskip\noindent
Thus, to find order-disorder dualities, one does not need to derive the (possibly complicated) rules for taking defects past field insertions, as in \eqref{eq:sig-defect-past-field}. Rather, it is enough to know the behaviour of the defect operators under composition.

\section{Orbifolds}

Denote by $G$ the group formed by the group-like defects of a rational CFT. The group $G$ automatically comes with further data, namely a class $[\psi]$ in $H^3(G,\mathbb{C}^\times)$, the third cohomology of $G$ with values in $\mathbb{C}^\times$ (with trivial $G$-action). 
This holds because group-like defects are described by invertible objects in a tensor category \cite{tft3}, and their associator gives rise to a 3-cocycle \cite[App.\,E]{Moore:1988qv}.
This 3-cocycle can also be seen directly on the level of correlators by employing defect junctions in the following way \cite{Runkel:2008gr}. For each pair $g,h \,{\in}\, G$ pick a non-zero topological junction field $\varphi_{g,h}$ for a 3-fold junction with defects $g$ and $h$ oriented towards the junction, and defect $g \,{\cdot}\, h$ oriented away from it. The space of such junction fields is one-dimensional \cite{Frohlich:2006ch}. Similarly, for three group-like defects $g,h,k \,{\in}\, G$, the space of topological junction fields for the four-fold junction with three in-coming defects labelled $g$, $h$, $k$, and one out-going defect $g \,{\cdot}\, h \,{\cdot}\, k$ is one-dimensional, too. As a consequence, the following proportionality between two configurations of defect lines and topological junction fields holds: 
\begin{equation}\label{eq:psi-def}
\setlength{\unitlength}{1.2pt}
\raisebox{-20pt}{\begin{picture}(50,50)(0,0)
    \put(0,0){
      \put(0,0){\scalebox{1.2}{\includegraphics{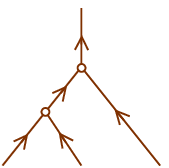}}}
      \put(0,0){
      \put(-4, 17){\scriptsize $\varphi_{g,h}$ } 
      \put(1, 29){\scriptsize $\varphi_{gh,k}$ } 
      \put(9, 4){\scriptsize $g$ } 
      \put(23, 4){\scriptsize $h$ } 
      \put(35, 4){\scriptsize $k$ } 
%      \put(19, 18){\scriptsize $gh$ } 
      \put(26, 40){\scriptsize $ghk$ } }
    }
\end{picture}}
 = ~ \psi(g,h,k) \, \cdot \!\!
\raisebox{-20pt}{\begin{picture}(50,50)(0,0)
    \put(0,0){
      \put(0,0){\scalebox{1.2}{\includegraphics{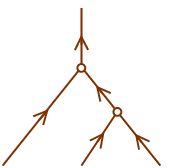}}}
      \put(0,0){
      \put(36, 17){\scriptsize $\varphi_{h,k}$ } 
      \put(26, 29){\scriptsize $\varphi_{g,hk}$ } 
      \put(7, 4){\scriptsize $g$ } 
      \put(20, 4){\scriptsize $h$ } 
      \put(34, 4){\scriptsize $k$ } 
%      \put(21, 16){\scriptsize $hk$ } 
      \put(26, 40){\scriptsize $ghk$ } }
    }
\setlength{\unitlength}{1pt}
\end{picture}}
\end{equation}
The pentagon identity for four-to-one junctions built from the $\varphi_{g,h}$ shows that
$\psi$ is a three-cocycle. Rescaling the junction fields $\varphi_{g,h}$ modifies $\psi$ by a coboundary. The basis-independent information is thus the class $[\psi] \in H^3(G,\mathbb{C}^\times)$.

The class $[\psi]$ provides an obstruction to orbifolding: one can only orbifold the CFT by subgroups $H \subseteq G$ such that $[\psi|_H]=1$. The origin of this obstruction can be understood as follows. Fix a subgroup $H \subseteq G$. To define the correlators of the $H$-orbifold theory, consider the superposition $Q = \sum_{h \in H} h$ of group-like defects. Then $\varphi = \sum_{g,h \in H} \varphi_{g,h}$ 
is a topological junction field for a two-to-one junction with all three legs labelled by $Q$. We will also need a topological junction field $\bar\varphi$ in the opposite direction, i.e.\ a one-to-two junction with all legs labelled $Q$:
\setlength{\unitlength}{1.2pt}
\begin{equation}\label{eq:phi-phibar}
\raisebox{-17pt}{\begin{picture}(25,35)(0,0)
    \put(0,0){
      \put(0,0){\scalebox{1.2}{\includegraphics{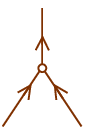}}}
      \put(0,0){
      \put(5, 17){\scriptsize $\varphi$ } 
      \put(-4, 5){\scriptsize $Q$ } 
      \put(22, 5){\scriptsize $Q$ } 
      \put(14, 27){\scriptsize $Q$ } }
    }
\end{picture}}
\qquad \text{and} \qquad 
\raisebox{-17pt}{\begin{picture}(25,35)(0,0)
\setlength{\unitlength}{1pt}
    \put(0,0){
      \put(0,0){\scalebox{1.2}{\includegraphics{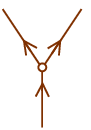}}}
      \put(0,0){
      \put(5, 15){\scriptsize $\bar\varphi$ } 
      \put(-4, 27){\scriptsize $Q$ } 
      \put(22, 27){\scriptsize $Q$ } 
      \put(14, 3){\scriptsize $Q$ } }
    }
\setlength{\unitlength}{1pt}
\end{picture}}
\end{equation}
Correlators of the orbifold model can be described as correlators of the un-orbifolded model with a sufficiently fine network of $Q$-defect lines inserted. These defect lines are joined at three-fold junctions and are oriented such that at each junction one can insert either $\varphi$ or $\bar\varphi$.

For example, on the torus one can take
\begin{equation}\label{eq:torus-example}
Z_\text{orb} =~ 
\raisebox{-28pt}{\begin{picture}(56,60)(0,0)
\put(0,0){\scalebox{1.2}{\includegraphics{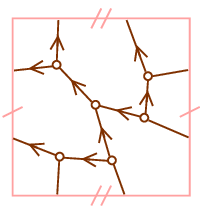}}}
\end{picture}
} ~~.
\end{equation}
For this description to be independent of the choice of defect network, it is clearly necessary that the defects and junction fields are topological. In addition, we must require invariance of the correlator under the following two local changes of the defect network:
\begin{equation}\label{eq:2d-Matveev}
\raisebox{-20pt}{\begin{picture}(45,45)(0,0)
\put(0,0){\scalebox{1.2}{\includegraphics{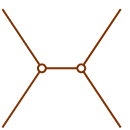}}}
\end{picture}}
\longleftrightarrow 
\raisebox{-20pt}{\begin{picture}(45,45)(0,0)
\put(0,0){\scalebox{1.2}{\includegraphics{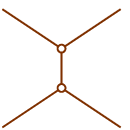}}}
\end{picture}}
\qquad \text{and}\qquad 
\raisebox{-20pt}{\begin{picture}(22,45)(0,0)
\put(0,0){\scalebox{1.2}{\includegraphics{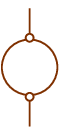}}}
\end{picture}}
~\longleftrightarrow ~
\raisebox{-20pt}{\begin{picture}(5,45)(0,0)
\put(0,0){\scalebox{1.2}{\includegraphics{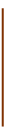}}}
\end{picture}}
\end{equation}
A defect network with three-valent vertices is dual to a triangulation of the surface, and the two local moves \eqref{eq:2d-Matveev} allow one to transform any dual triangulation into any other. The identities \eqref{eq:2d-Matveev} must hold for all allowed orientations of the defect lines. Expanding out the sum hidden in the superposition $Q$, we see that for three in-coming defects and one out-going defect, \eqref{eq:psi-def} implies invariance under the first move in \eqref{eq:2d-Matveev}, provided that $\psi(g,h,k)\,{=}\,1$ for all $g,h,k\,{\in}\, H$. A choice of $\varphi_{g,h}$ for which this is the case exists iff $[\psi|_H]=1$. One can convince oneself that then also $\bar\varphi$ exists such that \eqref{eq:2d-Matveev} holds for all allowed orientations \cite{tft3}.\footnote{One also needs that $D_g$ applied to the bulk vacuum $|0\rangle$ gives $|0\rangle$ rather than $-|0\rangle$; this is automatic in unitary models.}
 
In addition, any two possibilities of choosing $\varphi_{g,h}$ such that $\psi|_H\,{=}\,1$ are related by a 2-cocycle on $H$, and if we modify $\varphi_{g,h}$ by a factor that is a coboundary, this modification will cancel from \eqref{eq:torus-example}, and from all other correlators as well. 
One finds that different $H$-orbifolds are in one-to-one correspondence to $H^2(H,\mathbb{C}^\times)$ \cite{tft3}. This freedom is known as discrete torsion \cite{Vafa:1986wx,Kreuzer:1993tf}.

Recall that the crucial point in the construction of the orbifold correlators is the existence of a topological defect $Q$ with topological junction fields \eqref{eq:phi-phibar}, such that the modifications \eqref{eq:2d-Matveev} leave correlators invariant; it is not actually necessary for $Q$ to be a superposition of group-like defects. This opens the avenue for a generalisation of the orbifold construction by allowing more general topological defects $Q$, and thereby leads to the following result:

\medskip\noindent\emph{Every rational CFT that has identical left and right chiral symmetry $V$, that is well-defined on (oriented, closed) surfaces of genus 0 and 1, and that has a unique vacuum and non-degenerate 2-point correlator is a generalised orbifold of the Cardy case for $V$ (for which in particular the modular invariant torus partition function is given by charge-conjugation).}

\medskip\noindent
To see this, first note that in the TFT formalism \cite{tft1,tft5} a rational CFT with left and right chiral symmetry $V$ is described by a special symmetric Frobenius algebra in the category of $V$-modules. This Frobenius algebra provides us with the defect $Q$ and the vertices $\varphi$ and $\bar\varphi$. Comparing the construction of the correlator in terms of a Frobenius algebra \cite{tft5} and in the presence of a defect network \cite{Frohlich:2006ch}, one finds that they are exactly identical. Finally, every rational CFT satisfying the above conditions can indeed be described by a Frobenius algebra \cite{Fjelstad:2006aw,Kong:2008ci}.

Actually, one does not need to use the Cardy case as a starting point. Rather, every rational CFT satisfying the above conditions can be obtained from any other such CFT as a generalised orbifold. In the language of \cite{tft1,tft5,Frohlich:2006ch}, if CFT($A$) and CFT($B$) denote CFTs described by the special symmetric Frobenius algebras $A$ and $B$, respectively, then CFT($B$) can be obtained as a generalised orbifold from CFT($A$) via the defect\footnote{
$Q$ is an $A$-bimodule with left and right action of $A$ given by $m_A$, and $\varphi = \mathrm{id}_A \otimes \big( m_B \circ [\mathrm{id}_B \otimes (\varepsilon_A \circ m_A) \otimes \mathrm{id}_B]\big) \otimes \mathrm{id}_A$, where $m$ denotes the multiplication and $\varepsilon$ the counit, etc.}
$Q = A \otimes B \otimes A$.

To summarise: the study of topological defects and defect junctions leads to a generalisation of the concept of an orbifold which provides a uniform description of rational CFTs with left and right chiral symmetry $V$, including, in particular, exceptional modular invariants.

\vskip 3em

\newcommand\arxiv[2]      {\href{http://arXiv.org/abs/#1}{#2}}
\newcommand\doi[2]        {\href{http://dx.doi.org/#1}{#2}}
\newcommand\httpurl[2]    {\href{http://#1}{#2}}

\end{document}